\documentclass[aps,twocolumn,superscriptaddress,pre,nofootinbib,floatfix,usenatbib]{revtex4}
\usepackage{graphics}
\usepackage{graphicx,float}
\usepackage{epsfig}
\usepackage{amsmath}
\usepackage{amssymb}
\begin{document}
\title{A multi-messenger study of the total galactic high-energy neutrino emission}
\author{G. Pagliaroli}
\affiliation{Gran Sasso Science Institute, L'Aquila (AQ), Italy}
\affiliation{INFN-LNGS, L'Aquila (AQ), Italy}
\author{F.L. Villante}
\affiliation{Dipartimento di Scienze Fisiche e Chimiche, Universit\`a
  dell'Aquila,
L'Aquila (AQ), Italy}
\affiliation{INFN-LNGS, 
L'Aquila (AQ), Italy}

\begin{abstract}
A detailed multi-messenger study of the high-energy emission from the
Galactic plane is possible nowadays thanks to the observations
provided by gamma and neutrino telescopes and could be mandatory in
order to obtain a consistent scenario. 
We show the potential of this approach by using the
total gamma flux from the inner galactic
region measured by HESS at $1\,{\rm TeV}$  and in the longitude range $-75^\circ < l <
60^\circ$.
By comparing the observational data with the expected diffuse galactic emission, we highlight
the existence of an extended hot region of the gamma sky where the
cumulative sources contribution dominates over the diffuse component.
This region approximately coincides with the portion of the galactic plane from which a $\sim 2 \sigma$ excess 
of showers is observed in IceCube
high energy starting events.
In the assumption that hadronic mechanisms are responsible for the
observed gamma emission, we estimate the total galactic contribution 
(i.e. including both diffuse and the source components) to the IceCube
neutrino signal as a function of the spectral index and energy cutoff
of the sources, 
taking also into account the upper limit on a galactic component provided by Antares.    
\end{abstract}

\maketitle

\section{Introduction}
During the last years the IceCube collaboration provided the first clear evidence of an astrophysical flux of 
High Energy (HE) neutrinos \cite{Aartsen:2013jdh,Aartsen:2016xlq}. The
data collected in the energy range from 60 TeV to 10 PeV, 
called High Energy Starting Events (HESE), are consistent with an isotropic population of cosmic neutrinos having an energy distribution described by a power law with 
spectral index $2.92^{+0.33}_{-0.29}$ \cite{HESE6year}. At present,
no correlation of HESE arrival directions with the celestial positions
of known astrophysical sources has been found \cite{Aartsen:2017eiu}. 
The HESE data-set is dominated by showers events, characterised by poor angular
resolution, and it is mainly sensitive to the southern sky that
contains a large portion of the galactic plane and the galactic center.
On the other hand, the upward-going muons which originate from the northern sky
are better fitted by assuming a harder spectral index $2.13\pm0.13$ \cite{Aartsen:2016xlq},
so that a potential, despite still preliminary, tension exists between their
energy distribution and that of the HESE. 
This scenario could be consistent with the presence of a soft galactic component
which provides a relevant contribution in the southern sky in
addition to a harder, isotropic extragalactic neutrino flux
\cite{Denton:2017csz,Ahlers:2015moa,Neronov:2015osa,Palladino:2016zoe,Palladino:2016xsy}. 
However, no evidence for a galactic component is found in the most
recent dedicated analyses \cite{Adrian-Martinez:2016fei, Aartsen:2017ujz, Albert:2017oba}.

The existence of a non vanishing diffuse galactic contribution is
guaranteed by hadronic interactions of HE Cosmic Rays (CR) with the gas
contained in the galactic disk, through the production of charged
pions (and kaons) that subsequently decay to neutrinos. In addition
to this, HE neutrinos can be also produced by freshly accelerated
hadrons colliding with the ambient medium within or close to an acceleration site. 
Hadronic interactions produce a roughly equal number of charged and
neutral pions which decay to gamma rays. If photons are not absorbed by the
intervening medium, we thus expect that the HE neutrino sky is strongly
correlated with the HE gamma sky. 
This correlation provides us an handle to perform a detailed
multi-messenger study of the galactic plane. 
The results of HE gamma observatories can be combined with
the data collected by neutrino telescopes
in order to test their consistence in a coherent scenario.

In 2014, the H.E.S.S. Galactic Plane Survey \cite{Abramowski:2014vox} 
provided the first detailed observation of the large-scale
$\gamma$-ray emission in the inner region of the galactic plane at 
$E_\gamma\simeq 1$ TeV.
The measured flux represents the global emission from known and
unresolved sources, from diffuse $\gamma$-ray components and 
includes, in principle, the contributions of both hadronic and leptonic production 
mechanisms. 
%
In this paper, under the assumption that the observed HE photons are
mainly produced by hadronic interactions and that they are not
efficiently absorbed by the medium between the Earth and the
production point, we use the HESS data to estimate
the {\em total} HE neutrino flux from the galactic
disk, as a function of the neutrino energy and arrival direction.
This requires separating the sources contribution  
from the diffuse emission since the two components 
may have different spectral properties. 

By comparing the HESS data with theoretical predictions for 
the diffuse $\gamma-$ray flux,
we show that the source contribution dominates 
the $\gamma$ emission from the inner galactic region at 1 TeV and 
has a peculiar angular distribution that cannot be accounted by the
diffuse component. 
This permits us to identify a "hot'' extended region of the $\gamma-$ray sky 
which
could be also an important source of HE neutrinos. 
Interestingly, this region approximately coincides with the
portion of the galactic plane from which a 
$\sim 2 \sigma$ excess of showers is observed in the HESE IceCube data sample. 
We estimate the expected $\nu$ flux from this region
and we discuss the possibility to constrain neutrino sources emission
parameters
by considering the HESE
IceCube data and the upper limits on a possible galactic
component obtained by the  Antares neutrino telescope \cite{Adrian-Martinez:2016fei}.

 The plan of the paper is the following. In the first section we discuss the methodology and the assumptions used in this work.
 In Sec.~\ref{sec2} we calculate the neutrino and gamma diffuse galactic fluxes 
for different assumptions on the CR distribution.
 In Sec.~\ref{sec3} we compare our predictions for the diffuse gamma component with the total gamma flux observed by HESS. 
 The cumulative source contribution to galactic high-energy neutrinos is obtained in Sec.~\ref{sec4} as a function of source emission 
parameters. In Sec.~\ref{sec5} we summarize our results.

\section{Notations and methodology}
\label{sec1}
The total fluxes of HE neutrinos and gammas produced in our Galaxy can
be written as:
\begin{eqnarray}
\nonumber
\varphi_{\gamma, {\rm tot}} &=& 
\varphi_{\gamma, {\rm diff}} +
\varphi_{\gamma, {\rm S}} +
\varphi_{\gamma, {\rm IC}} \\
\varphi_{\nu,{\rm tot}} &=& 
\varphi_{\nu, {\rm diff}} +
\varphi_{\nu, {\rm S}} 
\end{eqnarray}
where $\varphi_{\gamma, {\rm diff}}$ ($\varphi_{\nu, {\rm diff}}$) is the diffuse gamma
(neutrino) flux produced by the interaction of CR with the gas
contained in the galactic disk, $\varphi_{\gamma, {\rm S}}$
($\varphi_{\nu, {\rm S}}$) is the gamma (neutrino) flux produced by
resolved and unresolved sources and $\varphi_{\gamma, {\rm IC}}$ is
the gamma flux produced through inverse compton by diffuse HE
electrons\footnote{We neglect possible contributions produced by DM annihilation
or decay and/or the possible
production of neutrinos and gammas in the galactic halo, being interested only in components that trace the galactic disk.}.  
With the term ``sources'', we refer here to  
all the contributions produced within or close to an acceleration
site by freshly accelerated particles that potentially have (a
part from cut-off effects) harder spectra than the diffuse component,
including thus resolved and unresolved 
objects.

 Our goal is to estimate the total neutrino flux 
$\varphi_{\nu,{\rm   tot}}$ from the galactic disk, as a function of the neutrino
 energy and arrival direction.
%
In order to do this, we calculate the
 diffuse gamma and neutrino flux as it described in the next section.
We then compare the diffuse gamma component with 
the observational determinations $\varphi_{\gamma,{\rm   obs}}$ 
of the {\em total} gamma flux at 1$\,{\rm TeV}$ obtained by the HESS detector.
This allows us to obtain by subtraction the HE gamma flux produced by
sources according to: 
\begin{equation}
\varphi_{\gamma, {\rm S}} \simeq
\varphi_{\gamma, {\rm obs}} - \varphi_{\gamma,{\rm diff}} 
\label{sources}
\end{equation}
In the above expression, we assume as a working hypothesis
that IC due to diffuse HE electrons provides a negligible contribution to the observed signal.
This seems plausible considering that the galactic component
is observed as the excess from the galactic plane with respect 
to the flux observed at larger galactic latitudes, automatically 
suppressing contributions with latitudinal intensity
profiles which are significantly more extended than those
derived by the gas distribution.
In particular, the signal in HESS is obtained as the excess relative to 
the $\gamma$-ray emission at absolute latitudes $|b|\ge 1.2^\circ$ \cite{Abramowski:2014vox} ; 
it has been evaluated that this implies a $\sim 95\%$ reduction of the
celestial IC signal \cite{Abramowski:2014vox}, assuming that this can
be modeled by using the Fermi-LAT detected diffuse galactic emission,
the GALPROP propagation code and the interstellar radiation field
model as done in \cite{Ackermann:2012pya}.

Finally, the neutrino flux emitted by sources is estimated by taking
advantage of the gamma/neutrino connection implied by hadronic
interactions. We assume that the differential
gamma flux is:
\begin{equation}
\label{gammaStheo}
\varphi_{\gamma, {\rm S}}
= 
k_\gamma({\hat n}_\gamma) 
\left(\frac{E_\gamma}{\rm TeV}\right)^{-\alpha_\gamma}
\exp\left(-\sqrt{\frac{E_\gamma}{E_{\rm cut,\gamma}}}\right)
\end{equation}
where the normalization $k_\gamma$  is determined as a function of the 
observation direction ${\hat n}_{\gamma}$ by requiring that:
\begin{eqnarray}
\label{gammaS}
\varphi_{\gamma, {\rm obs}}( {\hat n}_\gamma) &-& 
\varphi_{\gamma,{\rm diff}} (E_{\rm obs},\, {\hat n}_\gamma) 
= \\
\nonumber
&=& k_\gamma({\hat n}_\gamma) \left(\frac{E_{\rm obs}}{\rm TeV}\right)^{-\alpha_\gamma}
\exp\left(-\sqrt{\frac{E_{\rm obs}}{E_{\rm cut,\gamma}}}\right)
\end{eqnarray}
and the HESS observation energy is $E_{\rm obs} = 1\,{\rm TeV}$. 
If the gamma flux produced by sources is due to
hadronic interactions through $\pi_0$ (and $\eta$ mesons) decays, 
then a comparable neutrino flux is produced by the same objects 
through charged pions and kaons decays. This can be expressed
as a function of the neutrino energy $E_\nu$ 
and arrival direction ${\hat n}_\nu$ according to:
\begin{equation}
\label{nuStheo}
\varphi_{\nu, {\rm S}}
= k_\nu({\hat n}_\nu)
\left(\frac{E_\nu}{\rm TeV}\right)^{-\alpha_\nu}
\exp\left(-\sqrt{\frac{E_\nu}{E_{\rm cut,\nu}}}\right)
\end{equation}
where the neutrino spectral index and energy cutoff are given by \cite{Kappes:2006fg}:
\begin{eqnarray}
\nonumber
\alpha_\nu&=&\alpha_\gamma\\
E_{\rm cut,\nu} &=& 0.59 \,E_{\rm cut,\gamma}
\label{gammanu}
\end{eqnarray}
while the normalization constant can be obtained by using
\begin{equation}
\label{knu}
k_\nu ({\hat n}_\nu) = 
\left(0.694-0.16\alpha_\gamma\right)\, k_\gamma ({\hat n}_\gamma = {\hat n}_\nu)
\end{equation}
from the observational determination of
the total gamma ray flux and the knowledge of the diffuse gamma ray
component, see Eq. (\ref{gammaS}).

Few comments to above equations are in order. 
The parameterizations (\ref{gammaStheo}) and (\ref{nuStheo}) are obtained
in \cite{Kappes:2006fg} by assuming that photons and neutrinos are
produced through hadronic interactions by a CR population whose
spectrum is well described by a power law with 
exponential cut-off. Here, we are considering the cumulative gamma
and neutrino fluxes which are potentially produced by multiple (resolved and
unresolved) sources. Hence, we are automatically assuming that the average
spectrum of primary nucleons in the different sources is sufficiently 
well described by this functional form. 
It is possible, in principle, to adopt more refined approaches
\cite{Villante:2008qg} that do not require specific parameterizations of
the photon and neutrino flux. 
However, in consideration of the still incomplete knowledge of the 
sky at energies $\sim 1\,{\rm TeV}$ or larger, we believe that it is
advisable not to overcomplicate the model and to express the high
energy neutrino and gamma emission in terms of two parameters,
i.e. the neutrino 
spectral index $\alpha_\nu$ and the cutoff $E_{\rm cut, \nu}$ (or,
equivalently, $\alpha_\gamma$ and $E_{\rm cut,\gamma}$), which
may depend in principle on the observation direction. 
Unless otherwise specified, we assume for simplicity that they can be
considered constant in selected regions of the sky.

Finally, Eqs.(\ref{gammanu},\ref{knu}) that connect the gamma and neutrino
fluxes are obtained in the assumption that photons are produced in
the different sources through hadronic mechanism, with negligible
contribution of leptonic processes, and that they are not absorbed by 
the material between the production and the observation point. 
In the absence of specific observational evidence against this
assumption, we take it as a working hypothesis with the goal of
understanding the possibility to prove/disprove it with present and 
future neutrino telescopes.

\section{The diffuse galactic components}
\label{sec2}
The diffuse gamma and neutrino fluxes produced by the interaction of CR
with the interstellar medium in the galactic plane
can be written as:
\begin{eqnarray}
\nonumber
& &\varphi_{i,\text{diff}} (E_i,\hat{n}_i) = A_i \left[\int_{E_i}^{\infty} d E \, \right.
\frac{\sigma(E)}{E}\; 
F_i\left(\frac{E_i}{E},E \right)\, \\
& &   \left. 
\int_{0}^{\infty}  dl \,\varphi_{\rm CR} (E,{\bf r}_{\odot}+ l \, \hat{n}_i )\,
n_{\rm H} ({\bf r}_{\odot}+l \, \hat{n}_i ) \right],
\label{gammaflux}
\end{eqnarray}
where $i=\nu,\gamma$ stands for neutrinos and gamma respectively while
$E_i$ and $\hat{n}_i$ indicate the energy and arrival
direction of the considered particles. 
 The function $\varphi_{\rm CR} (E,{\bf r})$ represents the
differential CR flux, $n_{\rm H}({\bf r})$ is the gas density distribution and $ r_{\odot}=8.5$ kpc is the position of the Sun. 
The total inelastic cross section in nucleon-nucleon collision,
$\sigma(E)$, is given by:
\begin{equation}
\sigma(E) = 34.3 +1.88\ln(E/{\rm 1 TeV}) + 0.25 \ln(E/{\rm 1 TeV})^2
\;\; {\rm mb},
\nonumber
\end{equation}
where $E$ is the nucleon energy,  while the spectra $F_i \left(E_i/E,E  \right)$ of produced secondary particles 
are described (with 20\% accuracy) by the analytic formulas given in
\cite{Kelner:2008ke}. 
The constant $A_i$ is equal to $1$ for photons and $1/3$ for neutrinos.
The one flavour neutrino flux $\varphi_{\nu,\text{diff}}$ is indeed obtained by
summing over the production rates of $\nu_e$ and $\nu_\mu$ in the sources, i.e.:
\begin{equation}
 F_\nu \left(E_i/E,E  \right) \equiv F_{\nu_\mu} \left(E_\nu/E,E
 \right) + F_{\nu_e} \left(E_\nu/E,E  \right) \; ,
\end{equation}
and then assuming flavour equipartition at Earth, as it is expected with good
accuracy due to neutrino mixing, see e.g. \cite{Palladino:2015zua}.

Following our previous work \cite{Pagliaroli:2016lgg}, we consider different assumptions for the CR density in the
Galaxy that are intended to cover the large uncertainty in CR
propagation models. Namely, we assume that CR distribution is
homogenous in the Galaxy ({\em Case A}), that it follows the 
distribution of galactic CR sources ({\em Case  B}) and that 
it has a spectral index that depends on the galactocentric distance 
({\em Case C}).
These different assumptions permit us to relate the local determination of the CR flux,
$\varphi_{\rm CR,\odot} (E) $, to the CR flux in all the regions of the Galaxy where the
gas density is not negligible, according to:
\begin{equation}
\varphi_{\rm CR} (E,{\bf r})=
\begin{cases}
\varphi_{\rm CR,\odot} (E) \hspace{2cm}\text{\em{Case  A}}\\
\varphi_{\rm CR,\odot} (E)\, g({\bf r}) \hspace{1.345cm}\text{\em{Case B}}\\
\varphi_{\rm CR,\odot} (E)\,g({\bf r}) \, h(E,{\bf r}) \hspace{0.2cm}\text{\em{Case C}}.
\end{cases}
\end{equation}
For the CR flux at the Sun position, $\varphi_{\rm CR,\odot} (E)$, 
we consider the spectrum that was obtained in \cite{Ahlers:2015moa} 
by fitting the observational data of CREAM, KASCADE and KASCADE-Grande 
in the energy range $E\sim 1 -  10^6  \,{\rm TeV}$ in the assumption
that the dominant contributions to the nucleon flux are provided by 
${\rm H}$ and $^4{\rm He}$ nuclei. 

The function $g({\bf r})$ is proportional to the CR source density and 
it is obtained from the SNRs distribution of \cite{Green:2015isa}, as reported in Eq.(3.9) of \cite{Pagliaroli:2016lgg}.
The effect of this function is to increase the  CR density in {\em Case B} by a factor $\sim 4$   
at distances $r=2-3$ kpc from the galactic center with respect to the local value. 
The function 
\begin{equation}
h(E,{\bf r}) =\left( \frac{E}{\overline{E}} \right)^{\Delta({\bf r})}
\label{h}
\end{equation}
introduces a position-dependent variation $\Delta({\bf r})$ of the CR
spectral index in the {\em Case C}.
The pivot energy in eq.(\ref{h}) is taken as
$\overline{E}=20\;\rm{GeV}$,  
since it is observed \cite{Acero:2016qlg,Recchia:2016bnd}
that the integrated CR density above 20 GeV roughly follows 
the function $g({\bf r})$ (i.e. the SNR distribution). 
For our calculations, we take:
\begin{equation}
\Delta(r, z) = 0.3 \left(1-\frac{r}{r_{\odot}}\right)
\end{equation}
for $r\le r_{\odot}$, in galactic cylindrical coordinates, that 
is intended to reproduce the trend of the spectral index 
with $r$ observed by \cite{Acero:2016qlg} at 20 GeV.
This choice is equivalent to what is done by \cite{Gaggero:2017jts} in their 
phenomenological CR propagation model 
characterised by radially dependent transport properties.
Indeed, our calculations for {\em Case C} well reproduce 
the results of the KRA$_\gamma$ model both for HE photons 
and neutrinos in the energy range of interest for this analysis\footnote{
In this work, we do not use the high energy approximation adopted in
\cite{Pagliaroli:2016lgg}. We retain both the energy and position
dependence of the function $h(E,{\bf r})$ in such a way that our calculations 
are valid for energies larger than $\sim 10\,{\rm GeV}$ below which
the parameterizations of \cite{Kelner:2008ke} are no
more valid.}.

By following previous prescriptions, we can estimate the diffuse fluxes of high energy neutrinos and gammas   
at the different energies of interest and as functions of the galactic latitude $b$ and longitude $l$.
As described in our previous work \cite{Pagliaroli:2016lgg}, the diffuse fluxes are 
characterised by peculiar angular distributions. The maximal emission is always achieved for $l \simeq \pm 25^\circ$ and $b = 0^\circ$, 
but the fluxes may differ by large factors for $|l| \le 90^\circ$ in the three scenarios. 
To be quantitative, the diffuse gamma flux from the galactic
center at $E_\gamma = 1\,{\rm TeV}$ is larger by a factor $\sim 2$ and
$\sim 5$ in {\em Case B} and {\em C}
respectively, with respect to the value obtained in the assumption of
uniform CR density (i.e. {\em Case A}).

\section{The HESS extended hot region} 
\label{sec3}

\begin{figure}[t]
$$
\hspace{-0.7cm}
\includegraphics[width=0.5\textwidth]{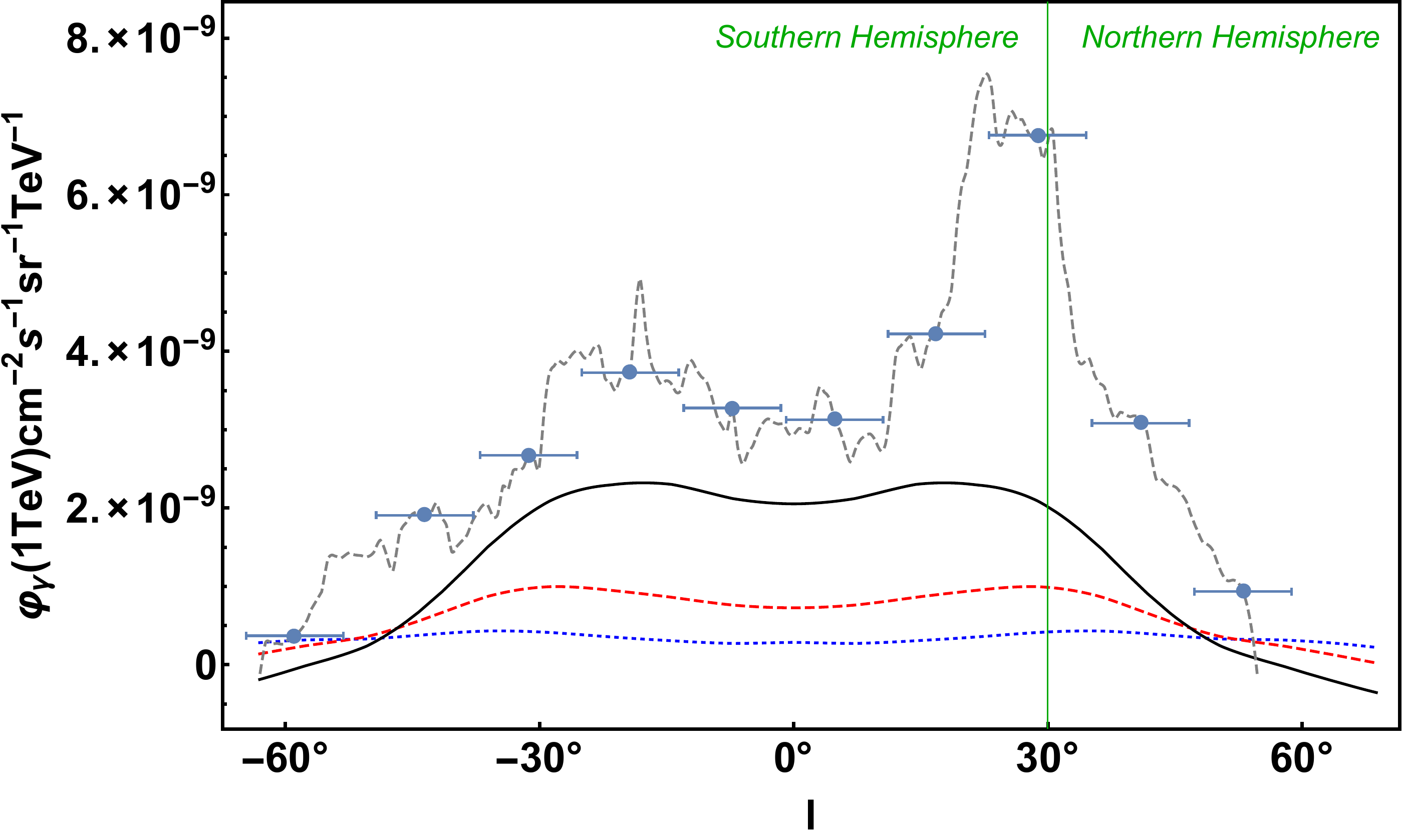}
$$
\caption{\em The gamma differential flux observed by HESS at  $E_\gamma=1$ TeV. 
The differential diffuse fluxes are also reported for a comparison
with a blue dotted line for {\em Case A} a red dashed line for {\em Case B} and a black line for {\em Case C}. 
The vertical line separates the Northern form the Southern Hemisphere.  
\label{fig0}}
\end{figure}

The first detailed observation of the large-scale $\gamma$-ray emission in the inner region 
of the galactic plane has been performed by the H.E.S.S. Galactic Plane Survey on 2014 \cite{Abramowski:2014vox}. 
HESS provides longitudinal and latitudinal profiles of the $\gamma$-ray emission 
at an energy $E_\gamma=1\,{\rm TeV}$, in the range 
of galactic longitude $-75^\circ < l < 60^\circ$ and galactic latitude $-2^\circ< b < 2^\circ$. 
The observed flux includes known sources, unresolved sources and diffuse $\gamma$-ray emission, so that, 
following eq.(\ref{sources}), it can be used to estimate the sources contribution in this region of the galactic plane.

The grey dashed line in Fig.\ref{fig0} shows the longitudinal profile of the 
total galactic emission observed by HESS. This is obtained by averaging the HESS data 
over an observation window $\Delta l \sim 15^\circ$, as emphasised by 
the horizontal error bar in the data points plotted in the Figure.
The re-binning of HESS data
is done for several reasons; first, we are interested in the cumulative emission 
from a given region of the sky without necessity of distinguishing between the different 
(resolved and unresolved) sources in each angular bin; second, the re-binning procedure avoids large fluctuations 
thus making visually clear the excess in each region of the sky with respect of the diffuse gamma 
expectations; third, our goal is to estimate the total galactic signal in IceCube HESE dataset, 
which is dominated by showers with an average $15^\circ$ angular resolution, thus not 
requiring a particularly detailed map of the galactic plane emission.

In order to compare the HESS data with the diffuse gamma fluxes
calculated in the previous section, we have to apply to our predictions the same background reduction procedure 
performed by the HESS collaboration and described in \cite{Abramowski:2014vox}.
For each considered case, we thus calculate the excess along the
galactic plane (i.e. in the region $|b|<1.2^\circ$) with respect to
the average emission in the region with $1.2^\circ <|b|<2^\circ$.
The resulting differential fluxes are reported as a function of $l$ in Fig.\ref{fig0} with a blue dotted line for {\em Case A}, a red dashed line for {\em Case B}
and a black line for {\em Case C}. 
It's evident from this plot that the gamma emission from the galactic plane 
is dominated at $E_\gamma=1\,{\rm TeV}$ by the sources contribution, 
even when the largest prediction for the diffuse component is considered. 
Namely, $\varphi_{\gamma,S}$ accounts for $89\%$, $76\%$ and $50\%$ of the
total observed gamma flux in {\em Case A}, {\em Case B} and {\em Case C}, respectively. 
Moreover the source contribution has a distinctive angular distribution; in all cases 
$\sim 50\%$ of this emission is concentrated in a broad peak with $11^\circ < l <57^\circ$ 
and  $|b|<2^\circ$. 
In the following, we refer to this specific range of galactic coordinates as the
Extended Hot Region (EHR) of the gamma sky. As shown in Fig.\ref{fig0}, 
only a fraction of this region (not containing the maximum of the expected emission) 
is contained in the Northern sky.

\begin{figure}[h]
\includegraphics[width=0.5\textwidth]{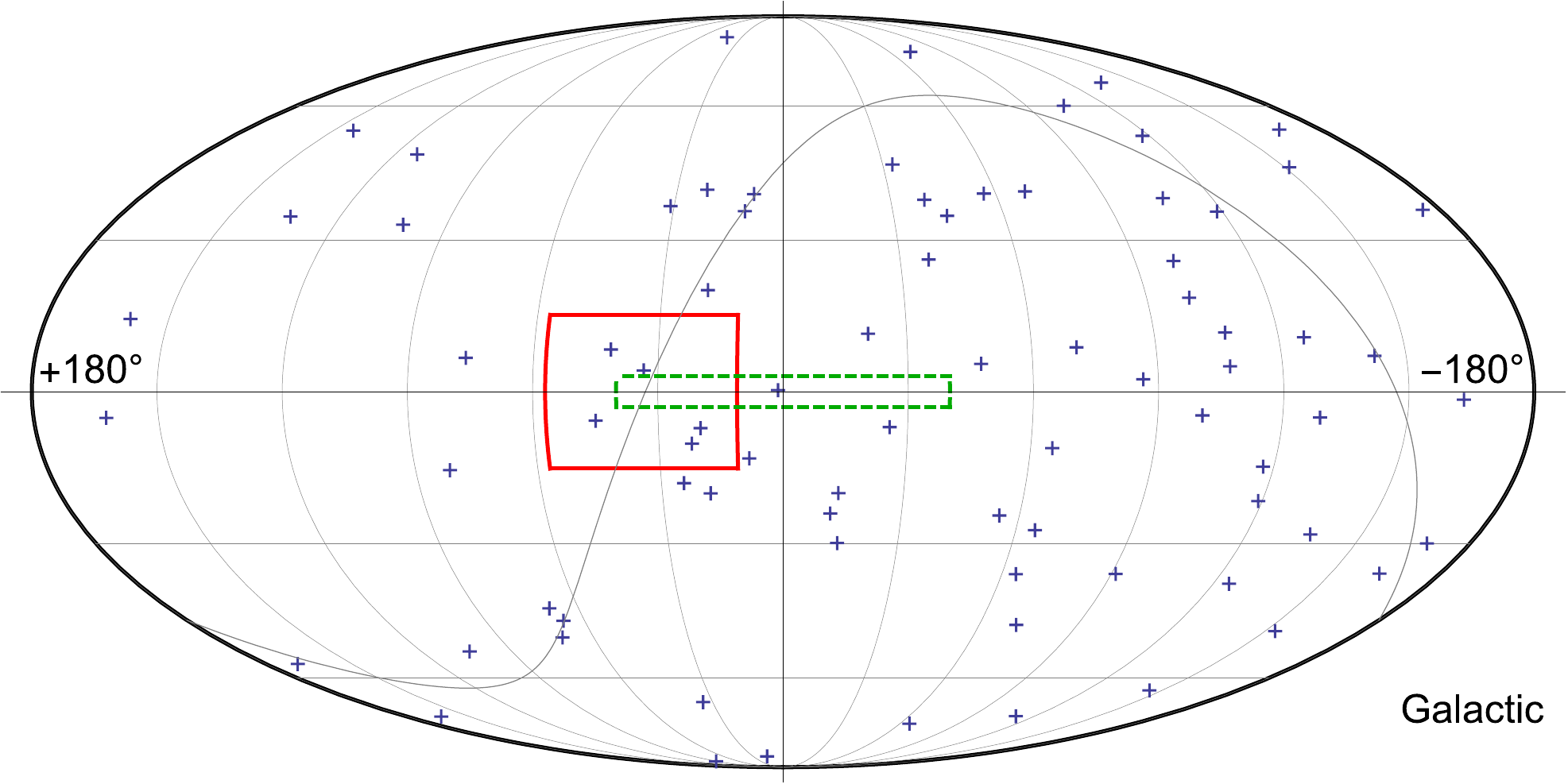}
\caption{\em The galactic distribution of the 6 years HESE collected by IceCube (blue crosses). The red box corresponds to the observation window defined in the text to probe the EHR of intense gamma emission from HESS. The green dashed box 
is the region considered by Antares to obtain the upper limit \cite{Adrian-Martinez:2016fei} on galactic neutrino emission. The grey line represents the equatorial plane.
\label{HESE6}}
\end{figure}

\section{The total galactic neutrino emission}
\label{sec4}

The IceCube detector probes the inner galactic region by using the HESE data set, firstly described in \cite{Aartsen:2013jdh}, 
that now includes $80$ events collected during 2078 days of data taking \cite{HESE6year}. 
These events are compatible with an isotropic best-fit flux
$E_\nu^{-2}\varphi_{\nu, \rm iso}=2.46\pm0.8\times10^{-8} (E_\nu/100 \rm{TeV})^{-0.92} \,\rm{GeV} \,cm^{-2}\, s^{-1}\, sr^{-1}$. 
 It is interesting to investigate whether an excess of HESE 
 exists from a region of the sky compatible with the EHR of gamma ray emission.
Since the HESE data set is mainly composed by showers, characterised by an average angular uncertainty of $\sim 15^\circ$, we define 
the observation window $|b|<15^\circ$ and $11^\circ<l<57^\circ$ that
corresponds to the red box in Fig.\ref{HESE6}. 
\begin{figure}[h]
\includegraphics[width=0.45\textwidth]{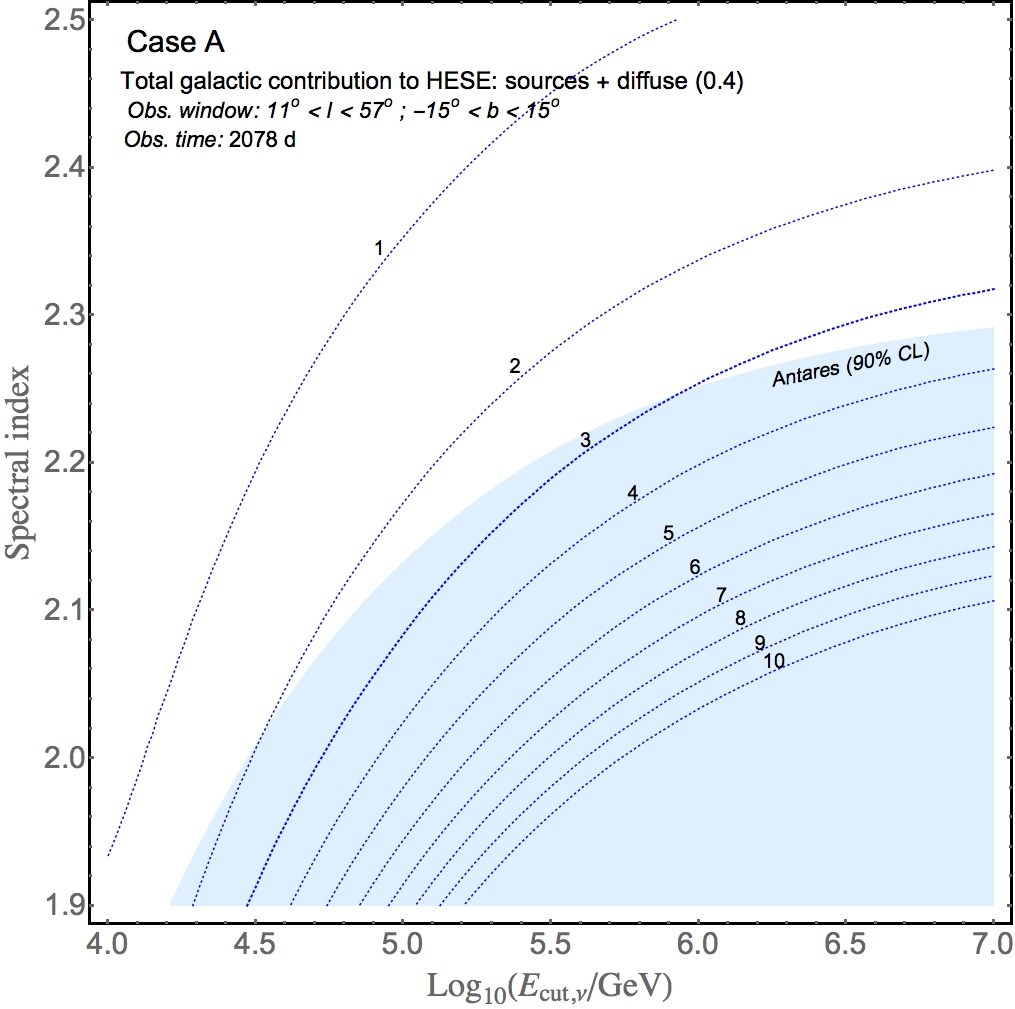}
\includegraphics[width=0.45\textwidth]{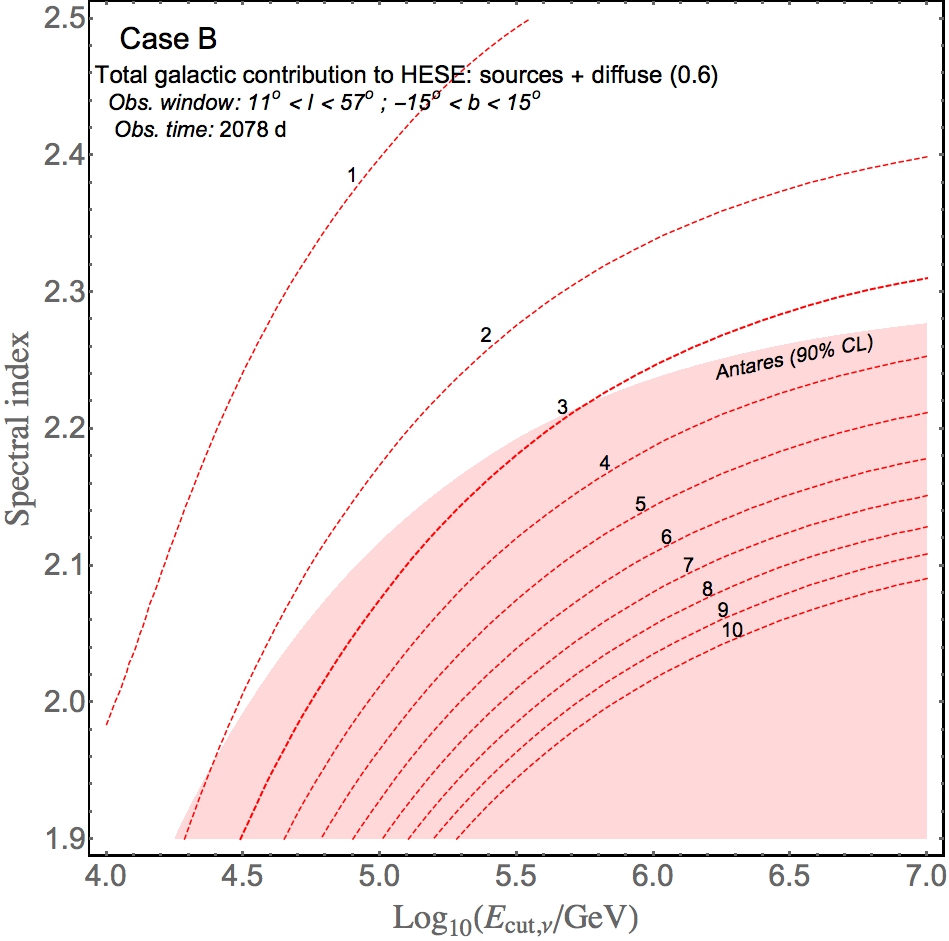}
\includegraphics[width=0.45\textwidth]{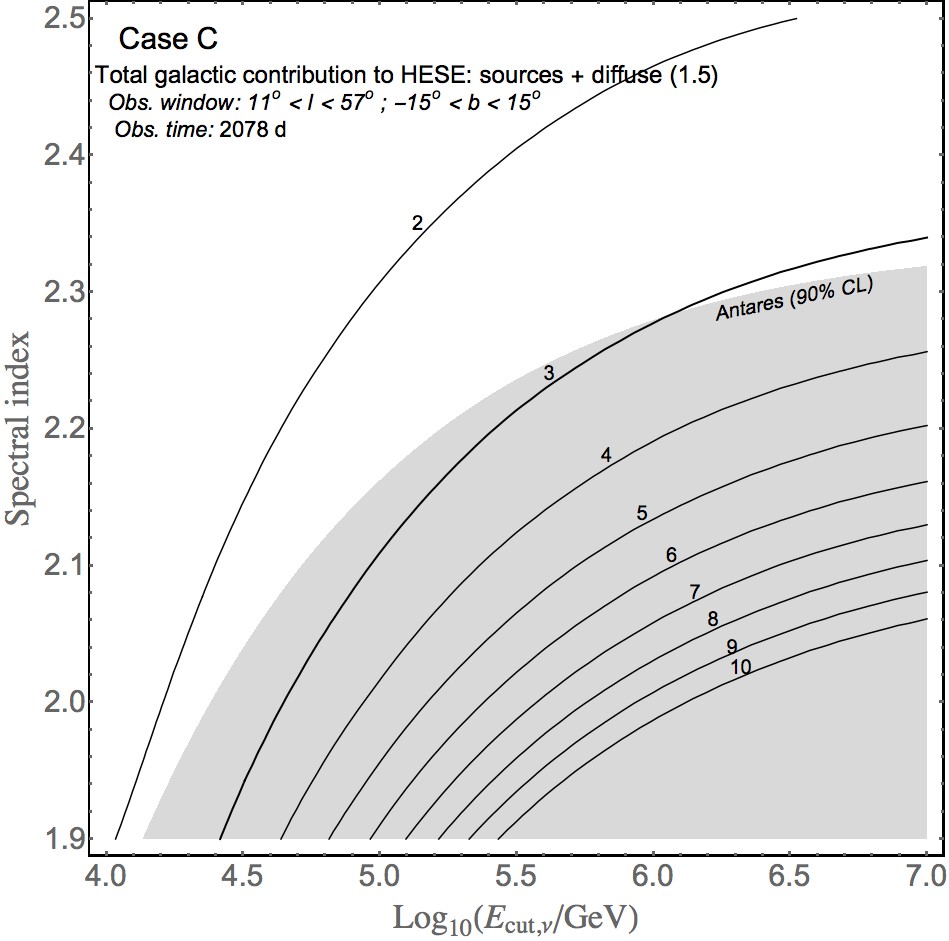}
\caption{\em 
The number of showers produced by the total (diffuse +
sources) galactic neutrino flux in IceCube in the indicated observation window for 
{\em Case A}  (upper panel),  
{\em Case B} (middle panel) 
and {\em Case C}  (lower panel).
The shaded areas show the regions of the plane 
$(E_{{\rm cut},\nu},\,\alpha_\nu)$
excluded by Antares.   
\label{fig1}}
\end{figure}

In the selected observation window, $N_{\rm Sh, obs}=5$ shower events (and no tracks) are observed; all of them have a relatively low reconstructed energy 
$E_{\rm dep}\sim 30 \,{\rm TeV}$, except for 
one event with  $E_{\rm dep}\sim 400 \,{\rm TeV}$. 
The number of observed showers should be compared with expectations from the atmospheric neutrino background\footnote{We do not include the atmospheric muon background because this mainly contributes to track events which are not observed in the considered observation
window.} and from the isotropic neutrino flux $\varphi_{\nu, \rm iso}$ which explains the bulk of IceCube HESE data.
The atmospheric neutrino background accounts for $\sim 15.6$ events (showers + tracks) in the whole sky. 
The number of showers in the considered observation window can be estimated by using the angular distribution of atmospheric neutrino events 
for $E_{\rm dep} > 30\,{\rm TeV}$  reported in Fig.5 of the Supplementary Materials of \cite{Aartsen:2014gkd} and by considering that showers are 
expected to be 31\% of the total events.
As a final result, we obtain $N_{\rm Sh, atmo} \sim 0.3$ 
from this background source, i.e. much lower than the observed number.

The contribution of astrophysical components (either extragalactic or galactic) is estimated by considering that the 
angular distribution of shower events can be calculated as (see \cite{Pagliaroli:2016lgg} for details):
\begin{eqnarray}
\label{Nsh}
\frac{dN_{\rm Sh}(\hat{n})} {d\Omega} &=&  T 
\int dE_\nu 
\int d\Omega_\nu\;
G_{\rm Sh}(\hat{n},\hat{n}_\nu)
\varphi_{\nu}(E_\nu,\hat{n}_\nu)  \\
\nonumber
& & \hspace{-1.2cm}
\times
\left[
A_{e}\left(E_\nu,\hat{n}_\nu\right) +
A_{\mu}\left(E_\nu,\hat{n}_\nu\right) (1-\eta) +
A_{\tau}\left(E_\nu,\hat{n}_\nu \right)  \right],
\end{eqnarray}
where $\hat{n}$ is the observation direction, $T$ is the observation time,
$A_{i}\left(E_\nu,\hat{n}_\nu\right)$ are the effective areas for the
HESE data sample \cite{Aree} 
and the parameter $\eta=0.8$ gives the probability that a muonic
neutrino produces a track event \cite{Palladino:2015zua}.
The function $G_{\rm Sh}$ is the showers angular resolution, i.e. 
\begin{equation}
G_{\rm Sh}(\hat{n},\hat{n}_\nu) = \frac{m}{2\pi \delta n_{\rm Sh}^2}\exp
\left(-\frac{1-c}{\delta n_{\rm Sh}^2}\right)
\end{equation}
where the parameter $m$ is a normalisation factor,
$c \equiv \cos \theta = \hat{n} \,\hat{n}_\nu$ describes the 
angle between the true ($\hat{n}_\nu$) and reconstructed ($\hat{n}$) neutrino direction
and the width $\delta
n_{\rm Sh}$ is calculated by requiring that $\theta\le
15^\circ$ at $68.3\%$ C.L. By using the above prescription, 
we estimate that the isotropic best fit astrophysical neutrino flux 
accounts for $N_{\rm Sh, iso}\sim 1.4$ showers 
in the considered observation window. 
In conclusion, we have an excess of 
$\Delta N_{\rm Sh}=N_{\rm Sh,obs}-N_{\rm Sh,atmo}-N_{\rm Sh,iso}\sim 3.3$ showers 
that corresponds to  $\sim 2\sigma$ fluctuation of the expected counting rate 
and that could be a potential indication in favour of a galactic contribution.

In view of the above results, we investigate whether the total galactic emission (i.e. diffuse + sources) can
provide a relevant contribution to the observed IceCube signal,
compatibly with the upper limit on the galactic component provided by
Antares.
The total galactic neutrino flux is estimated as a function of the
neutrino energy and arrival direction as explained in Sect.\ref{sec1}.
%
The neutrino angular distribution is fully determined by HESS
observational data, see eqs.(\ref{gammaS},\ref{knu}), while the energy
distribution depends on the spectral index $\alpha_\nu$ and the energy
cut-off $E_{{\rm cut},\nu}$ of the sources.
The coloured lines in Fig.\ref{fig1} correspond to 
a fixed number of shower events in IceCube produced by the {\em total} galactic component 
in the region $|b|<15^\circ$ and $11^\circ<l<57^\circ$
and during the observation time of 2078 days.
The three panels are obtained by calculating the diffuse neutrino
contribution as prescribed by {\em Case A}, {\em Case B} 
and {\em Case  C}, respectively. 
Not surprisingly, the total numbers of events produced in the three cases
are comparable, since the total gamma flux at 1 TeV is observationally
fixed by HESS (and implemented in our calculations).
However, the events are differently distributed among source and diffuse component
with a maximum (minimum) from the 
diffuse emission equal to 1.5 (0.4) in {\em Case C} ({\em Case A}).

The Antares neutrino telescope performed a detailed analysis of neutrino production from the central region of the 
galactic plane, i.e. $|l|<40^\circ$ and $|b|<3^\circ$, corresponding to the green dashed box in Fig.\ref{HESE6}, 
by using track-like events observed from 2007 to 2013 \cite{Adrian-Martinez:2016fei}.
No excess of events from the galactic ridge has been detected and $90\%$ upper limits on the galactic contribution 
averaged over the observation region have been set as a function of
the neutrino spectral index\footnote{We do not consider the more recent
bounds provided  by Antares \cite{Albert:2017oba} and IceCube \cite{Aartsen:2017ujz} 
because they are obtained by using the KRA-$\gamma$ model as template 
for galactic emission and cannot thus be applied  to constrain the total galactic emission (source + diffuse) which has a different angular and energy distribution.}.
For $\alpha_\nu=2.4$, the 90\% C.L.  upper limit at 100 TeV corresponds to $\varphi_\nu(100\,{\rm TeV}) = 2.0\times10^{-17}\, {\rm GeV}^{-1}\,{\rm cm}^{-2}\, {\rm s}^{-1}\, {\rm sr}^{-1}$
which is about a factor 2 larger than what predicted for the galactic
diffuse component in {\em Case C}.
If we assume that the sources emission parameters are
approximately constant in the EHR and in the Antares observation
window, we can implement the Antares bound in the plane 
$(E_{{\rm cut},\nu}, \,\alpha_\nu)$, excluding the shaded areas shown 
in the three panels of Fig.\ref{fig1}.

We see that significant constraints are obtained from Antares 
for $\alpha_\nu\le 2.3$; e.g. the possibility of a spectral index $\alpha_\nu =2.0$
 and cutoff energy $E_{\nu}\ge 30\,{\rm TeV}$ is excluded for all considered scenarios.
However, the Antares limit does not exclude the possibility that
galactic emission  could produce a non negligible event number in the IceCube HESE data
sample.

Indeed, up to $\sim 3$ shower events can be produced by galactic
neutrinos emitted from the EHR, compatibly with the Antares bound and
possibly accounting for a large fraction of the excess $\Delta N_{\rm
sh} = 3.3$ reported by IceCube.

\section{Summary}
\label{sec5}

In this paper, we perform a multi-messenger study of the total
galactic high-energy neutrino emission. 

By comparing the $\gamma$-ray observational data from H.E.S.S. Galactic
Plane Survey with the predicted diffuse galactic emission, we highlight
the existence of an extended hot region (EHR) of the gamma sky ($11^\circ<l<57^\circ$;
$-2^\circ<b<2^\circ$) where the cumulative sources contribution dominates over the diffuse component.
From the same portion of the galactic plane, we observe a $\sim 2 \sigma$ excess of shower events 
in the HESE IceCube data-set. 
Incidentally, the TeVCat catalogue \cite{TeVCat} contains about 20 unidentified $\gamma-$sources in this region
(most of them newly announced).

We investigate whether the total galactic emission (i.e. diffuse + sources) can
provide a relevant contribution to the observed HESE IceCube signal.
We show that the upper limit on the galactic contribution from Antares 
already provides significant constraints. 
However, it exists a region of the sources emission parameters (see
Fig. \ref{fig1}) that may explain the small excess of shower events from EHR 
observed by IceCube.
Dedicated analysis from Antares and IceCube to rule out this possibility
could be extremely interesting.

\acknowledgments{The authors acknowledge M. Cataldo, K. Egberts, C. Evoli and G. Morlino
for useful discussions.}

\bibliography{Bibliography1}

\end{document}